\documentclass{article}
\usepackage{spconf,amsmath,graphicx}

\usepackage[table]{xcolor}
\usepackage{color}
\usepackage{enumitem}
\usepackage{hyperref}
\usepackage{url}
\usepackage{breakurl}
\usepackage{multirow,makecell}
\usepackage{etoolbox,siunitx-v2}
\robustify\bfseries
\robustify\itshape
\usepackage{booktabs}
\usepackage{balance}
\usepackage{amssymb}
\usepackage{bm}
\usepackage{arydshln}
\usepackage{caption}
\usepackage{nicefrac}
\usepackage{cite}
\usepackage{cite}
\usepackage{bbm}
\usepackage{subcaption}
\usepackage{xspace}
\usepackage{scalerel}
\usepackage{lipsum}
\usepackage[scr=boondoxo]{mathalfa}

\def\setC{\mathscr{C}}

\def\setV{\mathscr{V}}

\def\ct{v}

\def\D{\mathcal{D}}
\def\E{\mathcal{E}}

\def\H{\mathcal{H}}
\def\L{\mathcal{L}}
\def\O{\mathcal{O}}

\def\T{\mathcal{T}}
\def\U{\mathcal{U}}
\def\one{\mathbbm{1}}

\usepackage{pifont}
\let\oldding\ding
\renewcommand{\ding}[2][1]{\scalebox{#1}{\oldding{#2}}}

\def\w{{{\theta}}}
\def\c{{c}}
\def\mix{{x}}
\def\es{\widehat{\mathbf{s}}}
\def\est{\widehat{{s}}_{\operatorname{T}}}
\def\eso{\widehat{{s}}_{\operatorname{O}}}
\def\s{\mathbf{s}}
\def\st{{s}_{\operatorname{T}}}
\def\so{{s}_{\operatorname{O}}}

\def\cit{OCT\xspace}
\def\citpp{OCT++\xspace}
\def\citineq{OCT\xspace}
\def\citppineq{OCT++\xspace}

\title{Optimal Condition Training for Target Source Separation}

\name{Efthymios Tzinis$^{1,2}$\thanks{Part of this work was performed as E.\ Tzinis was an intern at MERL. E.\ Tzinis was partially funded by the Google Ph.D.\ fellowship. E.\ Tzinis and P.\ Smaragdis were partially funded by NIFA grant \#2020-67021-32799. Code: \href{https://github.com/etzinis/optimal_condition_training}{https://github.com/etzinis/optimal\_condition\_training}}, 
Gordon Wichern$^1$, Paris Smaragdis$^{2}$, Jonathan Le Roux$^1$}
\address{
$^{1}$Mitsubishi Electric Research Laboratories (MERL), Cambridge, MA, USA \\
$^{2}$University of Illinois at Urbana-Champaign, Urbana, IL, USA
}

\begin{document}
\ninept
\maketitle
\setlength{\abovedisplayskip}{3pt}
\setlength{\belowdisplayskip}{3pt}
\begin{abstract}
  Recent research has shown remarkable performance in leveraging multiple extraneous conditional and non-mutually exclusive semantic concepts for sound source separation, allowing the flexibility to extract a given target source based on multiple different queries. In this work, we propose a new optimal condition training (OCT) method for single-channel target source separation, based on greedy parameter updates using the highest performing condition among equivalent conditions associated with a given target source. Our experiments show that the complementary information carried by the diverse semantic concepts significantly helps to disentangle and isolate sources of interest much more efficiently compared to single-conditioned models. Moreover, we propose a variation of OCT with condition refinement, in which an initial conditional vector is adapted to the given mixture and transformed to a more amenable representation for target source extraction. We showcase the effectiveness of OCT on diverse source separation experiments where it improves upon permutation invariant models with oracle assignment and obtains state-of-the-art performance in the more challenging task of text-based source separation, outperforming even dedicated text-only conditioned models.
\end{abstract}


\vspace{-.1cm}

\section{Introduction}
\vspace{-.1cm}
Humans possess the remarkable ability to isolate sounds from a noisy auditory input stimuli and associate them with objects and actions seamlessly. Auditory machine perception aims to mimic and even enhance this ability in a digitized manner, wherein the main challenge is to find an effective way to train models which are apt for the task of audio source separation.

Early works in deep-learning based audio source separation leveraged fundamental differences between the statistics of the sources of interest and those of other interfering sources in a mixture, making implicit assumptions on their semantic attributes. Thus, one could develop specialist models dedicating an output slot to recover only a given sound of interest, such as for speech enhancement \cite{xu2014experimental,Weninger2014RNN,erdogan2015psf,wang2018supervised} or instrument demixing \cite{jansson2017singing}. Eventually, more general training procedures such as deep clustering \cite{hershey2016deepclustering} and permutation invariant training (PIT) \cite{Isik2016Interspeech09,Yu2017PIT} took over the field, mainly because of their minimal a-priori assumptions on the types of sources. However, PIT's flexibility in training source separation networks does not come without a price, since PIT can neither solve the source alignment problem nor be used to explicitly specify the source of interest, and it suffers from instability problems~\cite{yang2020interrupted}. In contrast to semantically agnostic approaches, conditionally informed systems do not need to fix the order of the output sources and sometimes outperform PIT models \cite{le2015textInformed_SS, parekh2017motionInformedSS, schulze2019weaklyInformedSS}. Such works include models where an extra input conditional vector might carry information about speaker characteristics, musical instrument type, or general sound-class semantics, as proposed for speech \cite{delcroix2018single,ochiai2019unifiedInformedSpeakerExtraction,wang2019voicefilter,xiao2019single, zhuo_loc}, music \cite{Seetharaman2019ICASSPclass,Meseguer19CUNet,slizovskaia2021cunet}, and universal sound separation \cite{tzinis2020improving,ochiai2020listen,okamoto2021environmentalOnomatopoeia}.

Lately, there has been a resurgence of interest towards conditional separation models \cite{tzinis22_heterogeneous, ohishi2022conceptbeamTargetSpeechExtraction}, not only for boosting their performance but also to give the user more flexibility to query the model. In particular, heterogeneous speech separation \cite{tzinis22_heterogeneous} was recently proposed as a conditional source separation training procedure where non-mutually exclusive concepts are used to discriminate between a mixture's constituent sources. The resulting model not only can be queried using a diverse set of discriminative concepts (e.g., distance from the microphone, signal-level energy, spoken language, etc.), but also leverages the extra semantic information at training time to outperform PIT. Other follow-up works include single-conditioned models using a natural language description of the sources of interest \cite{liu2022separate_PlumbeyTextBasedSS} and/or encoded audio-snippet queries \cite{kilgour22_GoogleTextBasedSS}.

As the same target source may be queried using multiple equivalent conditions, in this work, we investigate whether for a given input mixture, an initial conditioning may be reformulated into a new conditioning that leads to better separation. As an intermediate step towards that goal, we first consider a system that focuses on reaching the best target extraction performance among all equivalent conditions for a given target, proposing a new training method, \cit, which performs a gradient step using the best performing conditional vector.
We then propose \citpp, which combines \cit with an on-the-fly conditional vector refinement module to reformulate, based on the input mixture, an initial query into a representation which can lead to better extraction of the target sources. We also extend the original heterogeneous training framework \cite{tzinis22_heterogeneous} to conduct experiments on the conditional separation of arbitrary sounds using more diverse and easy-to-use discriminatory semantic concepts such as \textit{text}, \textit{harmonicity}, \textit{energy}, and \textit{source order}.
Our experiments show that \cit yields a much higher upper bound for conditional separation based on the complementary semantic information of the diverse associated discriminative concepts surpassing all single-conditioned models and PIT. Moreover, \citpp yields state-of-the-art performance on text-based sound separation and surprisingly outperforms all dedicated 
text-based methods by a large margin.

 \vspace{-.2cm}
\section{Method}
\label{sec:method}
\vspace{-.1cm}

\begin{figure*}[h!]
    \centering
  \begin{subfigure}[h]{0.35\linewidth}
    \centering
      \includegraphics[width=1.\linewidth]{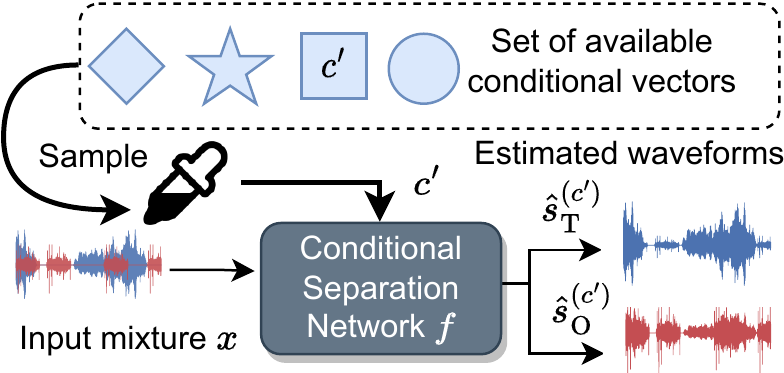}
      \caption{Heterogeneous condition training (HCT) \cite{tzinis22_heterogeneous}}
      \label{fig:methods:heterogeneous} 
     \end{subfigure} 
    \begin{subfigure}[h]{0.312\linewidth}
    \centering
      \includegraphics[width=1.\linewidth]{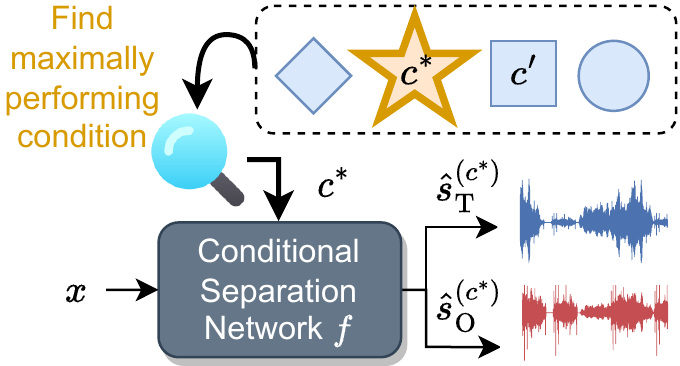}
      \caption{\cit training (ours)}
      \label{fig:methods:cit} 
     \end{subfigure} 
     \begin{subfigure}[h]{0.32\linewidth}
    \centering
      \includegraphics[width=1.\linewidth]{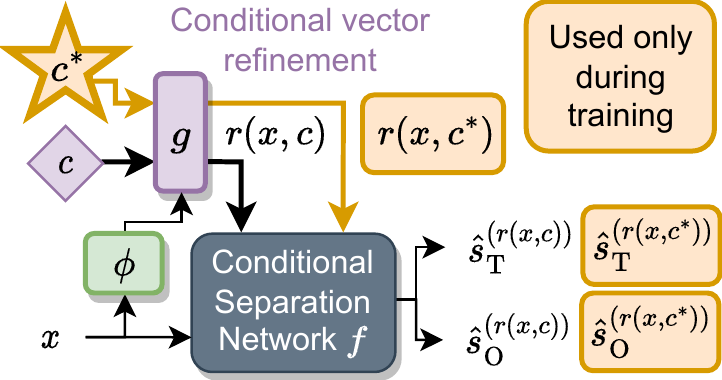}
      \caption{\citpp training (ours)}
      \label{fig:methods:citpp} 
     \end{subfigure} 
     \vspace{-.1cm}
     \caption{Different conditional separation training procedures for a given input mixture $\mix$. \ref{fig:methods:heterogeneous}: a heterogeneous condition vector $c'$ (associated with the target waveform $\st$) is sampled at random and a gradient step is performed. \ref{fig:methods:cit}: all available conditions are first evaluated and the error corresponding to the maximally performing conditional vector $c^*$ is backpropagated. \ref{fig:methods:citpp}: an initial condition vector of interest $c$ is first converted to a more amenable representation $r(x,c)$ using the trainable mappings $\phi$ and $g$, and the parameters are updated based on a regular \cit gradient update as well as the backpropagated errors from the regular path (black).}
    \label{fig:methods}
 \vspace{-.4cm}
\end{figure*}

We formulate the problem of conditional source separation as follows. Given an input mixture $\mix$ consisting of the sum $\mix = {\textstyle \sum}_{i=1}^{M} s_i$ of $M$ sources $\s=(s_1,\dots,s_M)$, we consider a target waveform $\st = {\textstyle \sum}_{j \in \mathcal{A}} s_j$ corresponding to a (potentially empty) subset $\mathcal{A} \subseteq \{1, \dots, M\}$ of target sources which can be described as associated with a condition $\ct$. Expressing $\ct$ as a conditional vector $\c=\c(\ct)$, we aim to train a model $f$ with parameters $\w_f$ which outputs estimates for both the target submix $\st$ and the non-target (``other'') submix $\so={\textstyle \sum}_{j \not\in \mathcal{A}} s_j$ of the mixed input sources $\s$:  
\begin{equation}
\label{eq:basic_estimation}
    \begin{gathered}
    \est^{(\c)}, \eso^{(\c)} = f(\mix, \c; \w_f).
    \end{gathered}
\end{equation}

The condition $\ct$ could be any discriminative concept which is associated with semantic characteristics of the target waveform $\st$. In this work, we consider the set of signal characteristics $\setC \!=\! \{\E, \H, \O, \T \}$, where $\E$ denotes the signal energy (with values low/high), $\H$ is the harmonicity of the target source (harmonic/percussive), $\O$ the order of appearance of the source in time (first/second), and $\T$ the text description of the target sound class(es) (e.g., a text embedding representing the words ``a dog barking'' given a mixture of sounds from an audio recording at a park).
Importantly, several conditions $\ct$ (and the corresponding conditional vector $\c(\ct)$) may be associated with the same target waveform $\st$. A schematic representation for all different conditional separation training methods discussed in this work is displayed in Fig. \ref{fig:methods}.

\vspace{-.1cm}
\subsection{Permutation invariant training (PIT)}
Usually, PIT \cite{Isik2016Interspeech09,Yu2017PIT} is employed for supervised training of unconditional source separation models by backpropagating the error using the best permutation $\pi \in \mathcal{P}_M$ of the set $\{1, \dots, M\}$ aligning the estimated sources $\es$ with the ground-truth sources $\s$ as shown next:
\begin{equation}
\label{eq:pit}
    \begin{aligned}
    \mathcal{L}_{\operatorname{PIT}}(\s, \bm{\theta}_f) = 
    \underset{\bm{\pi} \in \mathcal{P}_M }{\min} 
    [ {\textstyle \sum}_{i=1}^{M} \D(\es_{\pi(i)},s_i)  ], \enskip 
    \es = f(\mix, -; \w_f)
    \end{aligned}
\end{equation}
where $\D$ is any desired signal-level distance or loss used to penalize the reconstruction error between the estimates and their corresponding targets, and $-$ indicates the absence of conditioning. Notice that for the problem of target source separation, unconditional PIT models need to be considered in combination with a speaker selection scheme, since they do not solve the alignment problem of the estimated sources. Thus, we use the oracle permutation of sources, to study the upper bound of their separation performance.

\vspace{-.1cm}
\subsection{Heterogeneous condition training (HCT)}
The concept of heterogeneous condition separation, introduced in \cite{tzinis22_heterogeneous} for conditional speech separation, can be readily extended to general target sound source separation tasks. 
In essence, the model is fed with an input mixture $\mix$ as well as a one-hot encoded conditional vector $\c_{\operatorname{H}}(\ct) = \one[\ct] \in \{0, 1\}^{|\setV|}$ for the desired semantic concept $\ct$, where in \cite{tzinis22_heterogeneous} $\setV$ was a set of speaker discriminative concept values such as ``highest/lowest energy speaker'' or ``far/near field speaker.'' During training, a mixture $\mix$ is drawn or synthetically generated and an associated discriminative concept $\ct$ (corresponding to an encoded conditional vector $\c=\c(\ct)$) is drawn from a sampling prior $P(\ct)$ to form the desired target submix $\st$ containing all the sources $\s_i$ associated with $\ct$. The model tries to faithfully recover the target and non-target waveforms for $\ct$ by minimizing the following loss:
\begin{equation}
\label{eq:heterogeneous_loss}
    \begin{gathered}
    \L_{\operatorname{HCT}}
    (
    \st, \so, 
    \c;\theta_f) 
    = \D(\est^{(\c)}, \st) + \D(\eso^{(\c)}, \so),
    \end{gathered}
\end{equation}
where we explicitly stated $\st, \so, \c$ as parameters of $\L_{\operatorname{HCT}}$ to indicate that multiple combinations of conditions and targets may be considered. In \cite{tzinis22_heterogeneous}, it was shown that when the model is trained with multiple heterogeneous semantic conditional targets, an overall separation performance improvement can be achieved.

\vspace{-.1cm}
\subsection{Optimal condition training (OCT)}
As the same target waveform may be associated with multiple conditions, the question remains whether some conditions lead to better separation accuracy than others, and whether the system may benefit from modifying the conditioning vector based on the input, in other words to ``rephrase the query'' in light of the actual input. One reasonable goal to reach when modifying the conditioning would be the conditioning that obtains maximum performance for the given input mixture and target waveforms. A heterogeneous model may however need to balance its performance under multiple conditions, leading to suboptimal separation accuracy for the best conditioning, and thus ultimately for a system relying on modifying an original conditioning by replacing it or making it closer to the best one. Thus, we first consider training a model that solely focuses on optimizing performance for the maximally performing condition.

\cit follows a greedy approach in which instead of sampling a heterogeneous conditional vector $\c$ and training the separation system, several (potentially all) possible conditional vectors $\c' \in \mathcal{C}$ associated with the target waveform $\st$ are first evaluated, and we update the parameters of the network based on the condition that minimizes the overall error. Formally, we write the following loss function for updating the parameters of the conditional network as:
\begin{equation}
\label{eq:cit_loss}
    \begin{gathered}
    \c^* = \underset{\c' \in \mathcal{C} }{\operatorname{argmin}} \left[ \D(\est^{(\c')}, \st) + \D(\eso^{(\c')}, \so) \right], \\
    \L_{\operatorname{\citineq}}(\st, \so;\theta_f) = \L_{\operatorname{HCT}}(\st, \so,
    \c^*;\theta_f),
    \end{gathered}
\end{equation}
where $\c^*$ is the optimal condition (i.e., the one obtaining the smallest loss) for the input mixture $\mix$ and the target $\st$. We consider updating the model's parameters using conditional target vectors describing the ground-truth target submix $\st$ under various contexts sampled from the available signal characteristics $\setC \!=\! \{\E, \H, \O, \T \}$. For example, if one wants to train a conditional separation system based on text queries $\T$, there might be more effective ways to disentangle and isolate the same sources of interest based on complementary semantic information like the energy, the harmonicity, or the order of appearance of the sources. The evaluation of the ideal conditional target $\c^*$ is straightforward since we have access to the model $f$ and the ground-truth waveforms $\st$ and $\so$ during training. Of course, at inference time, one does not have access to the set of equivalent conditions to a given condition $\ct$, so focusing on improving only the optimal condition is not guaranteed to be a viable solution. This procedure was intended to serve as the basis for a method in which an auxiliary network refines an original condition by mapping it to the optimal equivalent one in light of the input mixture. One may in fact expect that focusing solely on maximally performing conditions, or in other words the easiest queries, may harm performance for other conditions. Surprisingly, the final conditional model learns how to associate the sources of interest with the corresponding semantic concepts and the overfitting problem can be easily avoided using an extra gradient update based on the condition of interest. \cit models can also perform better compared to dedicated systems trained and tested on the same input conditional information. 

\subsection{\citpp: \cit with embedding refinement}
Going a step further, there are cases where the input conditional information might not be informative enough by itself to lead to a conditioning vector that appropriately specifies the sources of interest, and one may hope to obtain an improved conditioning vector by letting the system look at both the input mixture and the original conditioning vector to output an improved conditioning vector. We thus consider introducing a learnable transformation $g(\cdot)$ of the conditional vector $\c$ to refine the conditional information so that it may be better utilized by the framework. For example, if the input mixture contains a guitar and a bass with different starting times, a query that corresponds to which instrument was played first ($\c_{\O}$: source order query) could be more informative than the textual description of the target musical instrument ($\c_{\T}$: text query). In that case, even if the user gives as an input \textit{retrieve the bass}, the learnable transformation $g$ could be used to map the less informative textual conditional input $\c_{\T}$ to something that resembles the ideal (oracle) conditional target $\c^* = \c_{\O}$. That transformation would in effect relieve the extraction network from making a difficult source selection and let it focus on the extraction. We let the learnable mapping $g$ take into account information about both the input mixture, via a time-invariant encoded representation $\phi(\mix; \w_{\phi})$, and the initial conditional target $\c$, computing the refined (or reassigned) conditional vector ${r}(\mix,\c)$ as:
\begin{equation}
\label{eq:reassignment}
    \begin{gathered}
    {r}(\mix,\c) = g(\operatorname{concat}(\phi(\mix; \w_{\phi}), \c); \w_g).
    \end{gathered}
\end{equation}

The final loss to be minimized combines the heterogeneous loss of Eq.\ \ref{eq:heterogeneous_loss} on the refined condition ${r}(\mix,\c)$ and the \cit loss of Eq.\ \ref{eq:cit_loss}, where $\c^*$ is the condition which leads to maximal performance after refinement. The loss is computed based on the refined counterpart ${r}(\mix,\c^*)$, as well as an extra regularizer term which aims to promote consistency at the conditional refinement mapping $g$ (e.g. steer the refined conditional target ${r}(\mix,\c)$ towards the ideal one ${r}(\mix,\c^*)$):
\begin{equation}
\label{eq:citpp_loss}
    \begin{gathered}
    \c^* = \underset{\c' \in \mathcal{C} }{\operatorname{argmin}} \left[ \mathcal{D}(\est^{({r}(\mix,\c'))}, \st) + \mathcal{D}(\eso^{({r}(\mix,\c'))}, \so) \right], \\
    \hspace{-1cm}\L_{{\operatorname{\citppineq}}}(\st, \so, \c;\bm{\theta})
    = \L_{\operatorname{HCT}}(\st, \so, {r}(\mix,\c);\theta_f)  \\
    + \L_{\operatorname{HCT}}(\st, \so, {r}(\mix,\c^*);\theta_f) + \| {r}(\mix,\c)  -  {r}(\mix,\c^*)\|^2, 
    \end{gathered}
\end{equation}
where the set of trainable parameters $\bm{\theta}=\{\w_f, \w_g, \w_{\phi}\}$ contains all the main network's $f$ parameters, the parameters of the conditional refinement mapping $g$, and the parameters of the mixture encoder $\phi$. In this case, the model tries to both optimize the separation performance of its estimate $\est$ as well as the reassignment mapping $g$ as it tries to make the conditional input vector look mostly like the highest performing conditional query after the transformation ${r}(\mix,\c^*)$.

\section{Experimental Framework}
\label{sec:exp_framework}

\subsection{Datasets}
\label{sec:exp_framework:datasets}
We extract the following three mixing datasets based on different portions of the FSD50K \cite{fsd50k_dataset} audio data collection, which consists of $200$ sound classes. Each training epoch consists of the on-the-fly generation of $20,000$ mixtures of $5$ s length, sampled at $8$ kHz and mixed at random input SNRs $\U [0, 2.5]$ dB with at least $80\%$ overlap (harder set) or $\U [0, 5]$ dB with at least $60\%$ overlap (easier set). The validation and test sets for each one of the following datasets are similarly generated, with $3,000$ and $5,000$ mixtures, respectively.

\noindent\textbf{Random super-classes}: We first randomly sample two distinct sound classes (out of the available $200$), then sample a representative source waveform for each class and mix them together.

\noindent\textbf{Different super-classes}: We select a subset of classes from the FSD50K ontology corresponding to six diverse and more challenging to separate super-classes of sounds, namely: \textit{Animal} (21 subclasses), \textit{Musical Instrument} (35 subclasses), \textit{Vehicle} (15 subclasses), \textit{Domestic \& Home Sounds} (26 subclasses), \textit{Speech} (5 subclasses) and \textit{Water Sounds} (6 subclasses). Each mixture contains two sound waveforms that belong to distinct super-classes.

\noindent\textbf{Same super-class}: Following the super-class definition from above, we force each mixture to consist of sources that belong to the same abstract category of sounds to test the ability of text-conditioned models in extremely challenging scenarios.

\subsection{Separation Model}
\label{sec:exp_framework:model}
We follow \cite{tzinis22_heterogeneous} and use the same conditional Sudo rm -rf model \cite{tzinis2022compute} with a trainable FiLM \cite{film} layer before each U-ConvBlock, with a mixture consistency layer at the output sources \cite{wisdom2019differentiableMixtureConsistency}, except that we here use only $B=8$ U-ConvBlocks since they were empirically found to be adequate for our universal conditional separation experiments. For the \citpp embedding refinement part, we use as $\phi$ the downsampling encoder part of one U-ConvBlock block with a similar configuration of $512$ intermediate channels and four $4$-strided depth-wise convolutional layers, and we reduce the time axis using a two-head attention pooling similar to \cite{tzinis2022audioscopev2}. The resulting vector is concatenated with the conditional vector $c$ and passed through $g$, which is a two-layer MLP with ReLU intermediate activations to form the refined conditional vector $r(\mix, c)$.

\subsection{Baseline systems}
\noindent\textbf{Text-based separation \cite{liu2022separate_PlumbeyTextBasedSS}}: We follow the previous state-of-the-art text-based source separation system proposed in \cite{liu2022separate_PlumbeyTextBasedSS} and use a pre-trained BERT \cite{devlin2019bert} encoding for the class of each sound. The final class encoding is computed after passing the first output token of the sequence model through a linear layer with a ReLU activation.

\noindent\textbf{Proposed text-based separation}: We also propose a stronger baseline for the text-based separation, wherein we replace the language model with a sentence-BERT model \cite{iandola2020squeezebert} and the first token with a mean average pooling operation and a trainable linear layer on top which better describes the linguistic information for shorter sentences like in audio-class based information (see results in Table \ref{table:final_text_results}).

\noindent\textbf{HCT \cite{tzinis22_heterogeneous}}: We train the system with equal sampling probability over all the available signal characteristics $\setC \!=\! \{\E, \H, \O, \T \}$.

\subsection{Training and evaluation details}
We train all models using the losses described in Sec. \ref{sec:method}. For the \cit text-based separation experiments we always perform a gradient update with both the text-query $c_{\T}$ and the best performing condition $c^*$ to avoid overfitting to the rest of the heterogeneous conditions. We use a batch size of $6$ and the Adam \cite{adam} optimizer with an initial learning rate of $10^{-3}$, halving it every $15$ epochs. 

We evaluate the source reconstruction fidelity at $110$ epochs, after empirically finding that all models had converged, using the mean scale-invariant signal-to-distortion ratio (SI-SDR) \cite{leroux2019sdr} between the estimate $\est$ and the ground-truth target $\st$. For the unconditional PIT oracle models, we measure the permutation invariant SI-SDR.

\section{Results}
\label{sec:results}

\subsection{Importance of the appropriate conditional vector}
\begin{figure}[t]
    \centering
  \begin{subfigure}[h]{\linewidth}
    \centering
      \includegraphics[width=1.\linewidth]{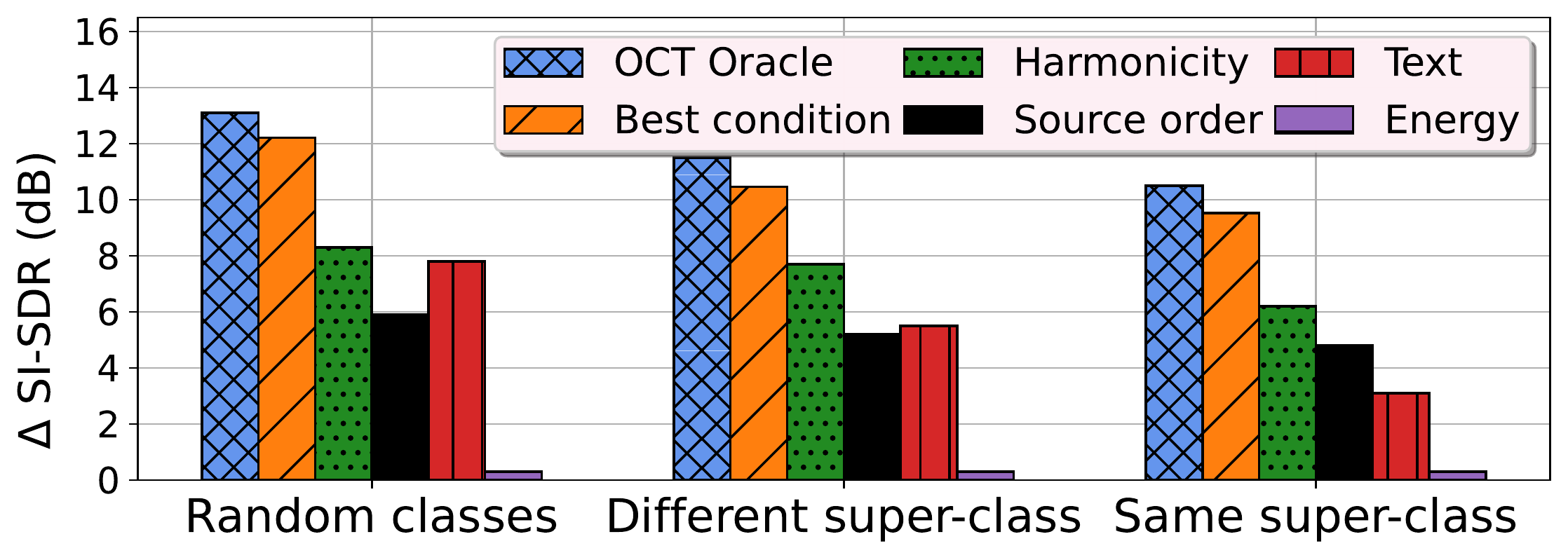}\vspace{-.3cm}
      \caption{Input-mixture SNR sampled from $\U [0, 2.5]$ dB.}
      \label{fig:diff_conditions_0_2.5} 
     \end{subfigure} \\
  \begin{subfigure}[h]{\linewidth}
    \includegraphics[width=1.0\linewidth]{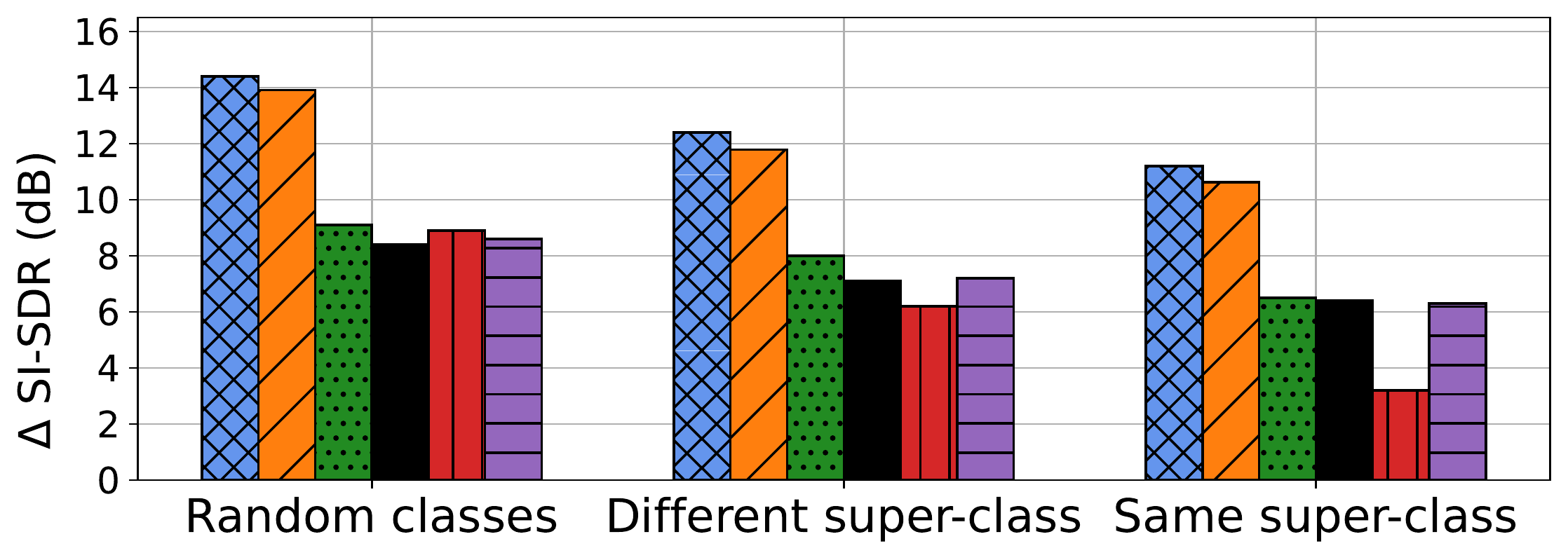}\vspace{-.3cm}
    \caption{Input-mixture SNR sampled from $\U [0, 5]$ dB.}\vspace{-.2cm}
      \label{fig:diff_conditions_0_5}
     \end{subfigure} 
     \caption{Mean test SI-SDR (dB) for target source separation with single-conditioned models using either \textit{harmonicity} ($\H$), \textit{source order} ($\O$), \textit{text} ($\T$), or \textit{signal-level energy} ($\E$) versus their oracle ensemble, which uses the best conditional vector, and an oracle \cit model trained and tested with the highest performing query.}\vspace{-.2cm}
    \label{fig:diff_conditions}
\end{figure}

In Fig.~\ref{fig:diff_conditions}, we show the performance of several single-condition models (trained to only handle a single type of query) and their oracle ensemble (where, for a given target, we select the query type leading to the best separation among all queries associated with the target) versus our proposed oracle \cit approach for target sound extraction. It is evident that several of the conditions fail dramatically on challenging data, while the best performing condition remains more robust, which indicates the importance of providing the right context for the task of target sound separation. For instance, the energy condition cannot be used when there is an ambiguity regarding the loudest source, as in cases where the input SNR is close to $0$ dB (see Fig.\ \ref{fig:diff_conditions_0_2.5}). Notably, the text-based condition, which is the most convenient to be used, performs poorly in the more challenging setups where the super-classes of sounds being mixed are similar or restricted, which enhances our belief that one needs to steer the conditional embedding vector towards the highest performing condition based on the given input mixture. Surprisingly, the \cit oracle model manages to perform better than the oracle best single-conditioned model which hints that integrating sound sources' semantic information through gradient-based updates can be an effective way for more robust source separation.

\subsection{\cit against state-of-the-art methods}
We choose text-based separation as our main benchmark since it is the most challenging condition and simultaneously the one that a user would likely use to describe the sources of interest. We measure the separation performance for the three universal separation datasets, as summarized in Table \ref{table:final_text_results}. It is evident that the oracle \cit method gives the best results even compared to the PIT oracle, which does not solve the estimated source alignment problem. We can thus assume that the complementary conditional information might be used to better disentangle the sources. Although our proposed single-conditioned text-based model surpasses the previous state-of-the-art text-based condition method \cite{liu2022separate_PlumbeyTextBasedSS} under all dataset configurations, it still performs poorly, especially for the harder to disentangle mixtures with input SNR in the $[0, 2.5]$dB range. Surprisingly, \cit, which was trained using the error signal from the best condition (which could be different from the text query), outperforms the dedicated text-based models, leveraging the complimentary information from the rest of the discriminative semantic concepts. \cit yields a significant improvement over heterogeneous training, which indicates that it is potentially a more efficient way of performing cross-semantic information training for source separation. Finally, our proposed embedding refinement method \citpp outperforms the previous state-of-the-art text-based separation method by $0.7$ to $2.6$ dB SI-SDR and yields a consistent improvement on top of the \cit by converting the conditional vector to a more amenable representation for text-based separation. We hypothesize that future work could provide much larger improvements by employing more sophisticated mixture encoders $\phi$ and refinement embedding maps $g$.

\begin{table}[]
\scriptsize
\centering
\caption{Mean test SI-SDR (dB) results for text-based sound source separation using mixing strategies with two levels of difficulty: input-SNRs $\U [0, 2.5]$dB and at least $80\%$ overlap (left) and $\U [0, 5]$dB with at least $60\%$ overlap (right). \textbf{Bolded} and \textit{Italic} numbers denote the best non-oracle and oracle models trained to perform text-based source separation, respectively.}\vspace{-.3cm}
\label{table:final_text_results}
\sisetup{table-number-alignment = center,
    detect-weight=true,
    detect-inline-weight=math,
    detect-all = true,
    mode=text,
    tight-spacing=true,
    round-mode=places,
    round-precision=1,
    table-format=2.1}
\setlength{\tabcolsep}{0.3em}
\resizebox{\linewidth}{!}
{%
\begin{tabular}[t]{l*{3}{S}|*{3}{S}
}
\toprule
\multicolumn{1}{l}{\multirow{3.6}{*}{\makecell{\textbf{Training method} \\ $^*$ Denotes our \\ implementation.}}}
& \multicolumn{3}{c|}{Input-SNR $\U [0, 2.5]$dB} & \multicolumn{3}{c}{Input-SNR $\U [0, 5]$dB}\\
\cmidrule(lr){2-4}\cmidrule(lr){5-7}
 & \multicolumn{3}{c|}{Super-classes in-mixture} & \multicolumn{3}{c}{Super-classes in-mixture}\\
 & \multicolumn{1}{c}{Random} & \multicolumn{1}{c}{Diff.} & \multicolumn{1}{c|}{Same} & \multicolumn{1}{c}{Random} & \multicolumn{1}{c}{Diff.} & \multicolumn{1}{c}{Same} \\
\midrule
Text only \cite{liu2022separate_PlumbeyTextBasedSS}$^*$ & 6.1 &3.9 & 2.2 & 8.6 &6.0 & 2.9 \\
Text only (ours) & 7.9 &5.6 & 3.1 & 9.0 &6.3 & 3.3 \\
HCT \cite{tzinis22_heterogeneous} & 6.8 &4.8 & 2.3 & 7.0 &4.3 & 2.4 \\
\midrule
(Proposed) \cit (No $\phi$ and $g$) & 8.4 & 6.0 & 3.3 & \bfseries 9.3 & 6.5 & 3.6 \\
(Proposed) \citpp & \bfseries 8.7 & \bfseries 6.2 & \bfseries 3.6 & \bfseries 9.3 & \bfseries 6.7 & \bfseries 3.7 \\
\midrule
\rowcolor[HTML]{E7E6E6}
(Oracle) \cit (No $\phi$ and $g$) &  13.1 &  \itshape 11.5 &  10.5 &  14.4 &  12.4 &  11.2 \\
\rowcolor[HTML]{E7E6E6}
(Oracle) \citpp & \itshape 13.2& \itshape 11.5 & \itshape 10.6 & \itshape 14.7 & \itshape 12.6 & \itshape 11.5 \\
\rowcolor[HTML]{E7E6E6}
(Oracle) PIT \cite{Yu2017PIT} & 12.4 & 10.7 & 9.8 &12.4 & 10.7 & 9.8 \\
\bottomrule
\end{tabular}
}
\vspace{-.4cm}
\end{table}

\section{Conclusion}
\label{sec:conclusion}
We have introduced a new training method for source separation which leverages the backpropagation of the optimal conditional vector signal. \cit outperforms all previous state-of-the-art single- and multi-condition (aka heterogeneous) training methods for the more challenging and easier-to-use text-based conditioning. Oracle \cit also outperforms unconditional models trained and evaluated with permutation invariance. \citpp enables further refinement by transformation of the conditional information vectors to a more amenable to separation form adapted to the input mixture. In the future, we aim to pair the proposed training methods with self-supervised approaches and explore in more detail the effectiveness of \cit.
\balance
\bibliographystyle{IEEEtran}
\bibliography{refs}

\end{document}